%% file: ms.tex
\documentclass[twocolumn]{aastex61}
\usepackage{amsmath}

\newcommand{\logg}{\ensuremath{\log g}}
\newcommand{\gflog}{\ensuremath{\log gf}}

\newcommand{\teff}{T$_{\rm eff}$}
\usepackage{graphicx}
\let\origsplitbox\splitbox
\let\splitbox\undefined
\usepackage{adjustbox}
\let\splitbox\origsplitbox

\usepackage{tablefootnote}
\usepackage{hyperref}
\usepackage[version=4]{mhchem}

\newcommand\aastex{AAS\TeX}

\received{\today}
\shorttitle{\aastex\ Elevated r-process enrichment in Gaia Sausage and Sequoia}
\shortauthors{Aguado et al.}
\begin{document}
   \title{Elevated r-process enrichment in Gaia Sausage and Sequoia\footnote{
Based on observations made with Very Large Telescope (VLT) at Paranal Observatory, Chile, under program 0104.B$-$0487(B)}
 }

%\correspondingauthor{David~S. Aguado}
\email{daguado@ast.cam.ac.uk}

\author[0000-0001-5200-3973]{David~S. Aguado}
\affiliation{Institute of Astronomy, University of Cambridge, Madingley Road, Cambridge CB3 0HA, UK \\}

\author[0000-0002-0038-9584]{Vasily Belokurov}
\affiliation{Institute of Astronomy, University of Cambridge, Madingley Road, Cambridge CB3 0HA, UK \\}

\author[0000-0002-5629-8876]{G.~C. Myeong}
\affiliation{Harvard  Smithsonian  Center  for  Astrophysics,  Cambridge, MA 02138, USA\\}
\affiliation{Institute of Astronomy, University of Cambridge, Madingley Road, Cambridge CB3 0HA, UK \\}

\author[0000-0002-5981-7360]{N. Wyn Evans}
\affiliation{Institute of Astronomy, University of Cambridge, Madingley Road, Cambridge CB3 0HA, UK \\}

\author[0000-0002-4343-0487]{Chiaki Kobayashi}
\affiliation{Centre for Astrophysics Research, Department of Physics, Astronomy and Mathematics, University of Hertfordshire, Hatfield, AL10 9AB, UK \\}

\author{Luca Sbordone}
\affiliation{ESO - European Southern Observatory, Alonso de Cordova 3107, Vitacura, Santiago, Chile\\}

\author[0000-0003-2481-4546]{Julio Chanam\'e}
\affiliation{Instituto de Astrofísica, Pontificia Universidad Católica de Chile, Av. Vicuña Mackenna 4860, 782-0436 Macul, Santiago, Chile\\}
\affiliation{Millennium Institute of Astrophysics, Av. Vicu\~{n}a Mackenna 4860, 782-0436 Macul, Santiago, Chile\\}

\author[0000-0002-4777-9934]{Camila Navarrete}
\affiliation{ESO - European Southern Observatory, Alonso de Cordova 3107, Vitacura, Santiago, Chile\\}
\affiliation{Millennium Institute of Astrophysics, Av. Vicu\~{n}a Mackenna 4860, 782-0436 Macul, Santiago, Chile\\}

\author[0000-0003-2644-135X]{Sergey E. Koposov}
\affiliation{Institute for Astronomy, University of Edinburgh, Royal Observatory, Blackford Hill, Edinburgh EH9 3HJ, UK\\}
\affiliation{Institute of Astronomy, University of Cambridge, Madingley Road, Cambridge CB3 0HA, UK \\}

%Harvard Smithsonian Center for Astrophysics, Cambridge, MA 02138, USA
%Centre for Astrophysics Research, Department of Physics, Astronomy and Mathematics, University of Hertfordshire, Hatfield, AL10 9AB, UK
%Millennium Institute of Astrophysics, Av. Vicuña Mackenna4860, 782-0436 Macul, Santiago, Chile
%Instituto de Astrofísica, Pontificia Universidad Católicade Chile, Av. Vicuña Mackenna 4860, 782-0436 Macul, Santiago, Chile
%ESO - European Southern Observatory, Alonso de Cordova 3107, Vitacura, Santiago, Chile

\begin{abstract}
The Gaia Sausage (GS) and the Sequoia represent the major accretion events that formed the stellar halo of the Milky Way. A detailed chemical study of these main building blocks provides a pristine view of the early steps of the Galaxy's assembly. We present the results of the analysis of the UVES high-resolution spectroscopic observations at the 8.2\,m VLT of nine Sausage/Sequoia members selected kinematically using \textit{Gaia} DR2. We season this set of measurements with archival data from Nissen \& Schuster (2011) and GALAH DR3 (2020). Here, we focus on the neutron-capture process by analysing Sr, Y, Ba and Eu behaviour.  We detect clear enhancement in Eu abundance ([Eu/Fe]$\sim0.6-0.7$) indicative of large prevalence of the $r$-process in the stellar $n$-capture makeup. We are also able to trace the evolution of the heavy element production across a wide range of metallicity. The barium to europium changes from a tight, flat sequence with [Ba/Eu]=-0.7 reflecting dominant contribution from exploding massive stars, to a clear upturn at higher iron abundances, betraying the onset of contamination from asymptotic giant branch (AGB) ejecta. Additionally, we discover two clear sequences in [Fe/H]$-$[Ba/Fe] plane likely caused by distinct levels of $s$-process pollution and mixing within the GS progenitor.
\end{abstract}



\section{Introduction} \label{sec:intro}

Detailed chemical evolution patterns from long-gone, primeval times can be gleaned today through spectroscopic studies of low-mass stars in dwarf galaxies. Many of the key elements, however, induce nothing but subtle imprints on stellar spectra, making abundance measurements in distant Galactic satellites hard, and thus leaving our view of the distinct enrichment pathways blurred. A new, powerful alternative is to look instead at the nearby (and therefore much brighter) stars deposited into the Galaxy together, as part of a past merger event \citep[e.g.][]{Roederer2010,agu20}. The trick therefore is to grasp which stars in the halo's hotchpotch belonged to the same progenitor. The flip side of the coin is that the story of the early mergers that built the Milky Way is hard to unravel, but luckily there is important evidence in the chemical abundances themselves.

\citet{niss10, Nissen11,nissen12} (hereafter NS) first identified two sequences in the local halo population from stellar abundances. There is a higher [$\alpha$/Fe] sequence, corresponding to high star formation rates and a lower [$\alpha$/Fe] sequence corresponding to slower enrichment; two sequences overlap considerably in iron abundance, but the high-$\alpha$ reaches higher metallicities. NS argued that the lower $\alpha$-sequence corresponds to populations accreted from dwarf galaxies.  This prediction was subsequently confirmed by data from the \textit{Gaia} Satellite, when \citet{belo18} identified the \lq\lq Gaia Sausage'' (GS) as the residue of a nearly head-on merger event $\sim$ 10 Gyrs ago~\citep[see also][as
well as \citet{Ev20} for a review of the history of the idea]{brook03,Helmi2018,Ha18,mye18a}. Subsequently, \citet{mye19} argued that the stars in the eccentric, highly retrograde halo came from an additional event termed the \lq\lq Sequoia''\citep[see also][]{Ma19}.

This has stimulated recent high resolution spectroscopic studies of small samples~\citep{matsuno20, rohan20,monty20,ve20, limberg20}, as well as large-scale medium resolution surveys, such as GALAH~\citep{galahdr3}. Here, we present results from our own sample of GS and Sequoia members, together with reanalyses of NS and GALAH data, with a focus on the neutron capture elements (Sr, Y, Ba and Eu).
\input{parameters2.tex}

\section{High-resolution spectroscopy}\label{analisis}

\subsection{UVES Target Selection and Observations} \label{sec:obs}
We observed a total of nine potential GS and Sequoia members selected from \textit{Gaia}'s Radial Velocity Spectrometer \citep[RVS,][]{gaia2018, gaia_rvs}. The orbital parameters were calculated for \textit{Gaia} DR2 stars with provided six-dimensional phase space information using the distance estimates of \citet{anders2019}. An axisymmetric potential model of the Milky Way \citep{mcmillan2017} has been adopted for the calculation. Nine stars with good visibility were selected based on the known kinematic characteristics of each halo component \citep[see e.g.,][]{belo18,mye18a,mye19,monty20}.

The selected sample was observed under the ESO program 0104.B-0487(B) with the UVES spectrograph \citep{dek00} on the 8.2\,m Kueyen Very Large Telescope at Cerro Paranal Observatory, in the Atacama desert, Chile. The used set-up was dichroic Dic \#1\,(390+580) with $1\farcs2$ slit and $1\times1$ binning with moon minimum angular distance equal to $90$\,deg  and a maximum airmass of $\sim1.4$. This setup provided a spectral coverage between 330 and 680~{nm} and nominal resolving power of R$\sim41,000$ for the blue part of the spectrum ($330-452$~{nm}) and R$\sim39,500$ for the red ($480-680$~{nm}). However, the seeing during observations spans  $0\farcs50-1\farcs45$ and therefore in some cases resolving power is higher up to 45,000.
Each target was visited once between 15th December 2019 and 17th March 2020 in queue mode. We aimed for relatively high signal-to-noise (S/N) ratio in order to detect weak metallic absorptions ($\sim$75 at 393\,nm). Table \ref{parameters} summarises observing parameters and obtained S/N.  The REFLEX environment was used to reduce the data within the ESO Common Pipeline Library.

\subsection{Stellar Parameters}

The radial velocity ($\mathrm{v_{rad}}$) was measured on each individual spectrum by using a cross correlation function over the \ion{Mg}{1}b triplet region ($\sim517$\,nm). The used template was a synthetic spectrum with similar stellar parameters computed with the SYNTHE code. To check possible $\mathrm{v_{rad}}$ variability, we compared our UVES measurements with those from Gaia's RVS. Values are shown in Table \ref{parameters} and no radial velocity variation could be detected at a level larger than $\sim 1$\,km/s. However, due to the low number of available measurements binarity behaviour should not be discarded at this stage. Once the spectra are heliocentric corrected, we normalize them using a running mean filter with a 300-pixel window. Then we perform a global analysis using the FERRE\footnote{{\tt FERRE} is available from http://github.com/callendeprieto/ferre} code \citep{alle06}. It allowed us to derive T$_{\rm eff}$, $\log \mathrm{g}$, [Fe/H] and [C/Fe] using the same grid of synthetic models as in \citet{agu20} computed by the ASSET code and assuming a fixed microturbulence of 2\,km s$^{-1}$. FERRE searches for the best fit interpolating between the nodes of the grid using the Boender-Timmer-Rinnoy Kan algorithm to minimize the $\chi^{2}$. Table~\ref{parameters} lists the stellar parameters and carbon abundance derived in this analysis.
\begin{figure*}
\begin{center}
{\includegraphics[width=145 mm]{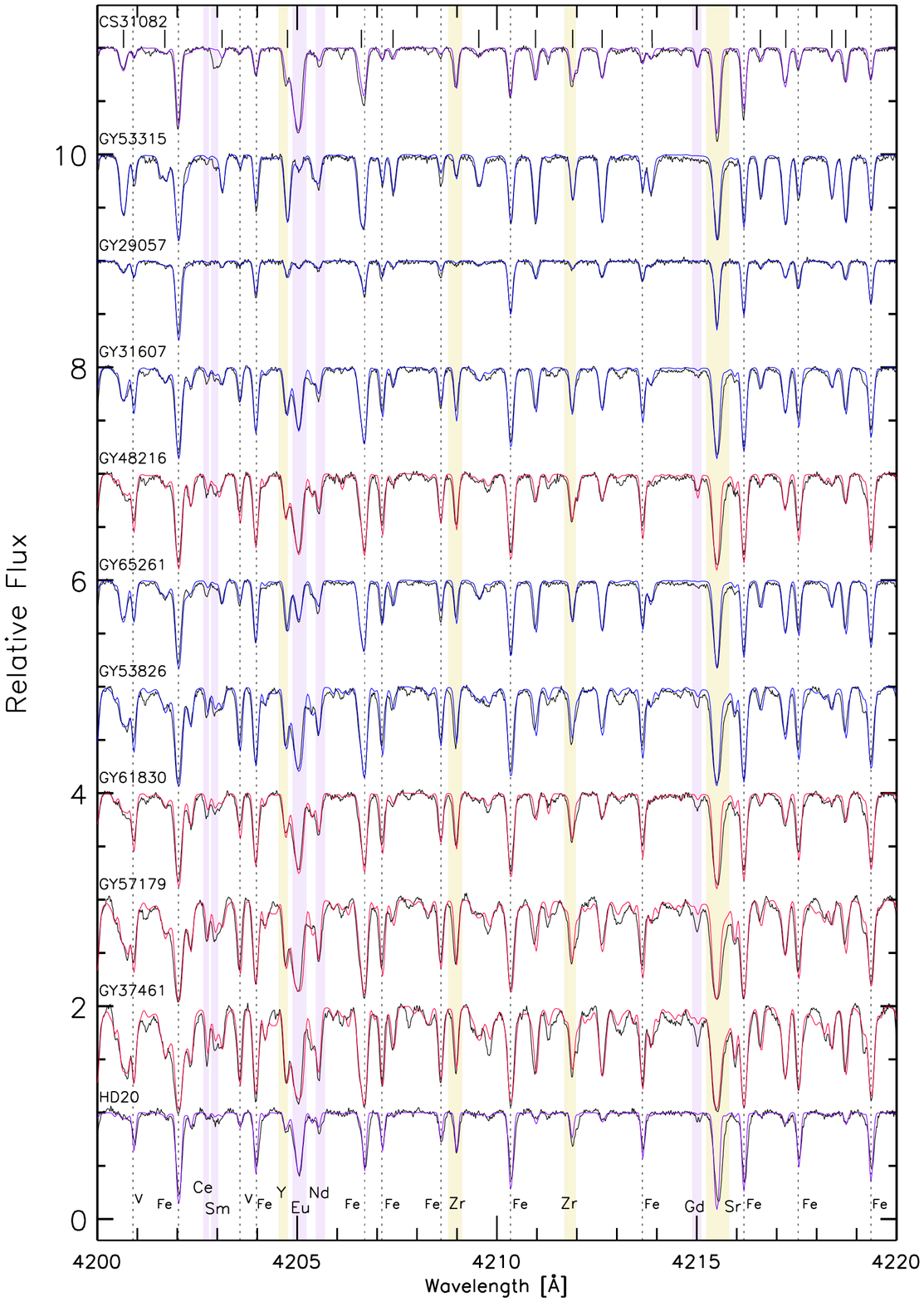}}
\end{center}
\caption{ A narrow region of the UVES spectra (4200\,\AA-4220\,\AA, black line) of our stellar GS and Sequoia sample sorted by metallicity and the best SYNTHE model (red for GS and blue for Sequoia). For comparison, the UVES spectra of two well known metal-poor Eu-rich stars with different metallicities are also shown with the best model in purple: HD 20 from \citet{barklem05} and CS 31082$-$001 from \citet{hill02}. Coloured areas show the main r-process (blue) and s-process ones (yellow) absorption features. The main iron-peak elements are labelled with a dashed line while short thick black lines in the top of the figure denote the strongest CH features.}
\label{spectra}
\end{figure*}

\subsection{Elemental Abundances}

For the purpose of this paper, we focus on n-capture elements (Sr, Y, Ba and Eu). The full chemical signature (including C and $\alpha$ abundances) of this GS sample deserves a separate analysis and will be discussed in detail in future work (Aguado et al., in preparation).
Deriving n-capture elemental abundances in metal-poor stars is always difficult due to the relatively weak lines available in the optical spectrum. With the exception of the strongest Sr and Ba lines, all other absorptions require high-resolution spectroscopy. The wide range of \teff\, ($\sim1700$\,K) implies large variations in the depths of the lines. Additionally, different evolutionary phases ($\logg \sim 1.2-4.5$) strongly affect the intensity of the weaker lines. The line list used was the most updated Castelli-Kurucz one from 2016\footnote{Available in http://wwwuser.oats.inaf.it/castelli/linelists.html}. We compute for each target a collection of synthetic spectra with the SYNTHE code \citep{kur05} based in ATLAS9 \citep{mez12} atmosphere models. These synthetic spectra include 0.1\,dex variation of each individual species.  Then we derive each abundance by interpolating between individual SYNTHE models and summarise the results in Table \ref{parameters}. The detected lines used in this work are 4077 and 4215\,\AA\,\, for \ion{Sr}{2}; 3600, 3710, 3774, and 4374\,\AA\, for \ion{Y}{2}; 4934, 6141, and 6497\,\AA\,\, for \ion{Ba}{2}; 4129, 4205, 6437 and 6645\,\AA\, for \ion{Eu}{2}. In the case of Sr where both lines are saturated in most of the spectra, we find systematically lower Sr abundance ($\sim 0.2$) in the redder (4215\,\AA) line which is also severely blended and often discarded. By using only the Sr 4077\,\AA\, line we recover compatible abundance from a similar well known star HD 20.  In Fig \ref{spectra}, the 4200$-$4220\,\AA\, region is shown with the best model for all the GS and Sequoia sample and two metal-poor Eu-rich stars, HD 20 \citep{barklem05} and CS 31082$-$001 \citep{hill02}. This blue area is particularly rich in n-process absorptions but also contains plenty of iron-peak, $\alpha-$ and CH$-$ lines.

The high r-process production detected here made the Eu determination specially challenging. In nature, Eu has only two stable isotopes \ce{^{151}_{63}Eu+} and \ce{^{153}_{63}Eu+} and a broad isotopic structure \citep{sne08}. We adopted the isotopic ratios of 47.8\% and 52.2\% respectively from \citet{lawler2001} that are in agreement with the chondrite ratio measured in the Solar System \citep{lod03}. However, the Eu hyperfine structure (HFS) has dramatic consequences for the shapes of the most intense lines. One cannot reproduce the wings of the lines without taking into account detailed HFS. Then, we used central wavelengths and \gflog\, from \citet{lawler2001}. The fact that the objects display high Eu over-abundance could make the strongest line go into the saturated growth curve regime, especially in the metal-rich end of our sample ($\rm [Fe/H]\sim-1.5$). Even in non-severely saturated lines, this could lead to an inaccurate measurement. For that reason, the weaker Eu line at 6645\,\AA, is commonly used to avoid this problem \citep[see, e.g.,][]{han18}. Some cases are also reported in the literature \citep[see, e.g.,][]{ryab99} in which the determined abundances from the red-weaker lines are slightly lower than those from the blue-stronger ones. This difference may be related to the NLTE effect affecting differently the blue and the red lines. Finally, some of the Eu lines such 4205\,\AA\, are severely blended with Carbon and other iron-peak elements \citep{lawler2001}. Therefore, we used i) the four Eu lines when available and non-saturated (GY31607 and GY65261), ii) the less blended blue line 4129\,\AA\, if the red lines the not detected (GY29057 and GY53315) which the most metal-poor ones), and iii) only the red ones (6437 and 6645\,\AA) when saturation happened (GY37461, GY48216, GY53826 GY57179, and GY61830).
 
\begin{figure}
\begin{center}
{\includegraphics[width=60 mm, angle=90]{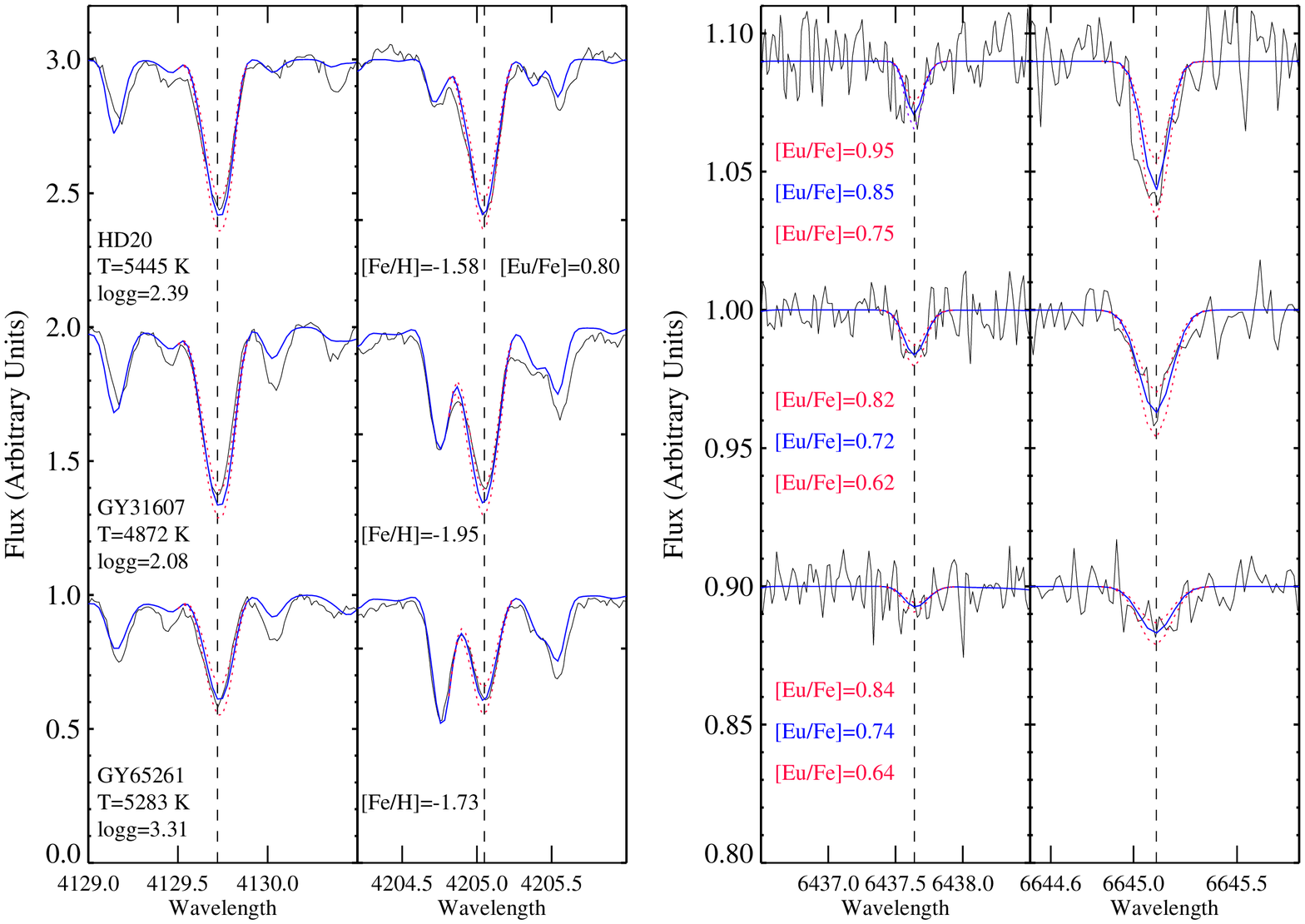}}
\end{center}
\caption{ Several portions of the UVES spectra for GY31607, GY65261, and the calibration star HD 20 (black lines), together with the best Eu fit (blue lines) and upper and lower limits (red lines). Stellar parameters for each star are also shown. Notice the scales are not the same in the left and right panels. See text for discussion. }
\label{europium}
\end{figure}

 The consistency of this methodology is shown in Fig. \ref{europium}, where we perform a comparative analysis with the well-known metal-poor Eu-rich HD 20 star.  Notice that the scale is not the same in the left and the right column of Fig. \ref{europium} -- they are different by a factor of $\sim10:1$. Following the procedure explained above, we derive the same Eu abundance as that presented in \citet{barklem05} within the errors. Moreover, in a more recent analysis at very high S/N, \citet{hanke20} found even closer Eu value and therefore validating the followed methodology. We also show detailed Eu analysis for GY31607 and GY65261 with compatible but sometimes different values from the four lines. Indeed, the use of different Eu lines could lead to slight inhomogeneities or offsets that we estimate to be about $\sim0.1$\,dex according to our comparative analysis presented in Fig. \ref{europium}.
 
 The 1D-LTE approach this work used has some impact on the derived n-process abundances. According to \citet{mash14}, NLTE corrections lead to a lower Ba but higher Eu in r-rich stars. That means the [Ba/Eu] ratios presented in Table \ref{parameters} could potentially decrease by $\sim0.1-0.2$\,dex. Furthermore, the authors in \citet{gall20} calculated for the first time 3D-NLTE corrections within $\lesssim0.05$\,dex for the Ba resonance lines included in this study.

\begin{figure*}
\begin{center}
{\includegraphics[width=77mm,angle=90,trim={ 6.8cm .cm .cm 0cm},clip]{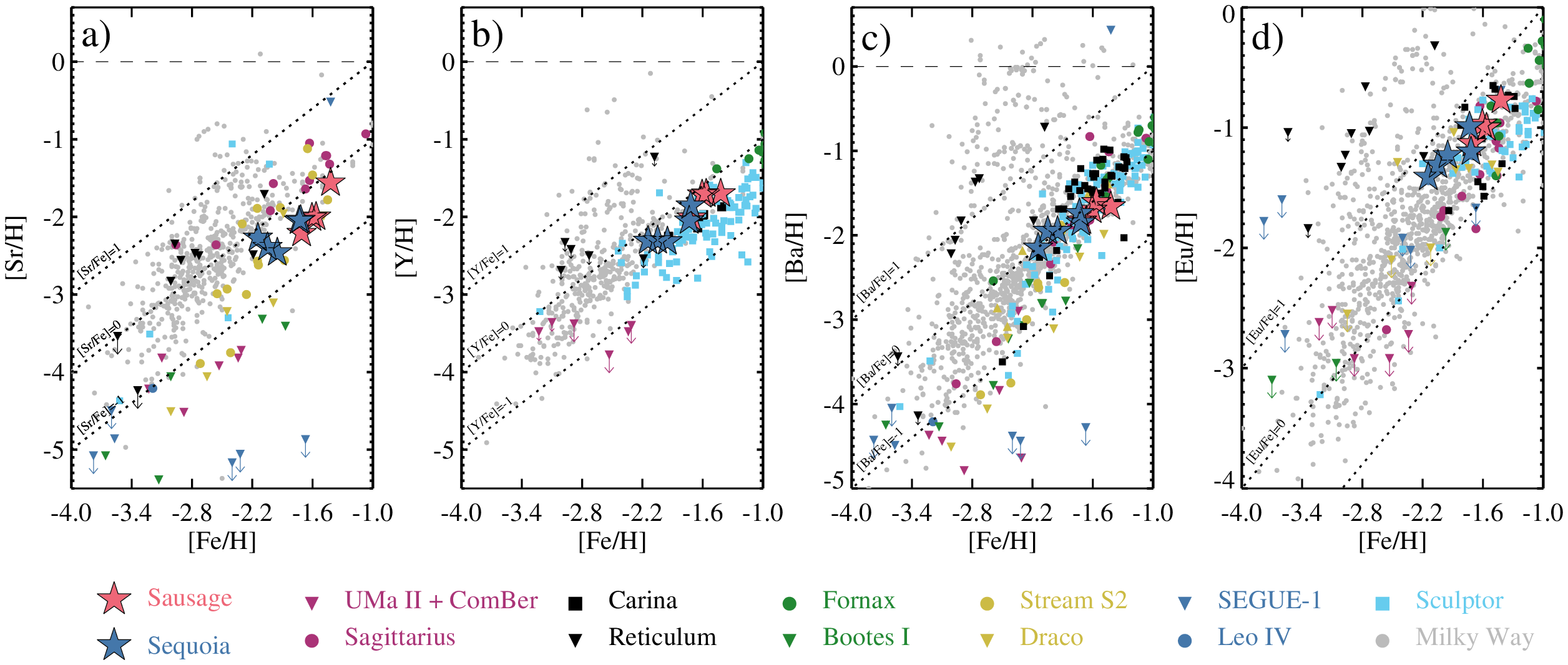}}
{\includegraphics[width=66mm,angle=90,trim={ 8.5cm .0cm 0.cm 0cm},clip]{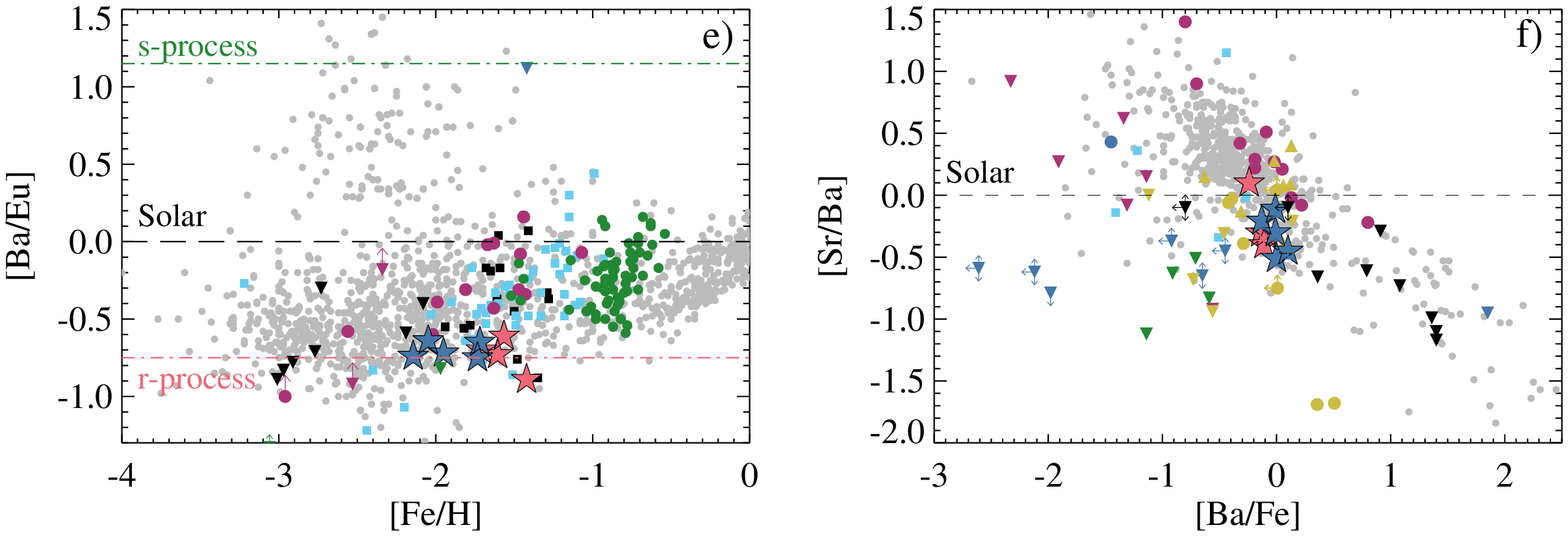}}
\end{center}
\caption{ N-capture abundances ratios derived in this work for GS and Sequoia. We also show elemental abundances from other ultra faint and classical dwarf galaxies from literature: Ursa Major II \citep[][]{freursa10},  SEGUE-1 \citep[][]{fresegue14}, Sculptor \citep[][]{ jabsculp15, hill19}, Coma Bernices \citep[][]{freursa10}, Sagittarius \citep{hansensag18}, Draco \citep[][]{cohendraco09, shedraco13}, Carina \citep[][]{venncarina12}, S2 \citep[][]{agu20}, Bootes \citep[][]{norrbootes10,laibootes11}, Leo IV \citep[][]{simonleo10}, Reticulum II \citep[][]{jireti16}, and Fornax \citep[][]{letfornax18}. For comparison we also show those of halo from JINA \citep{jina18}, SAGA \citep{saga08} and R-Process Alliance \citep{ezzeddine20}, and disk stars from \citet{bens14}.  In panel $e$ pure $s$- and $r$-process production from \citet{bis14} are shown in green and red dashed lines.
}
\label{elemental}
\end{figure*}

\begin{figure*}
\begin{center}
{\includegraphics[width=\textwidth, angle=0,trim={ 0 .0cm 0.cm 0cm}]{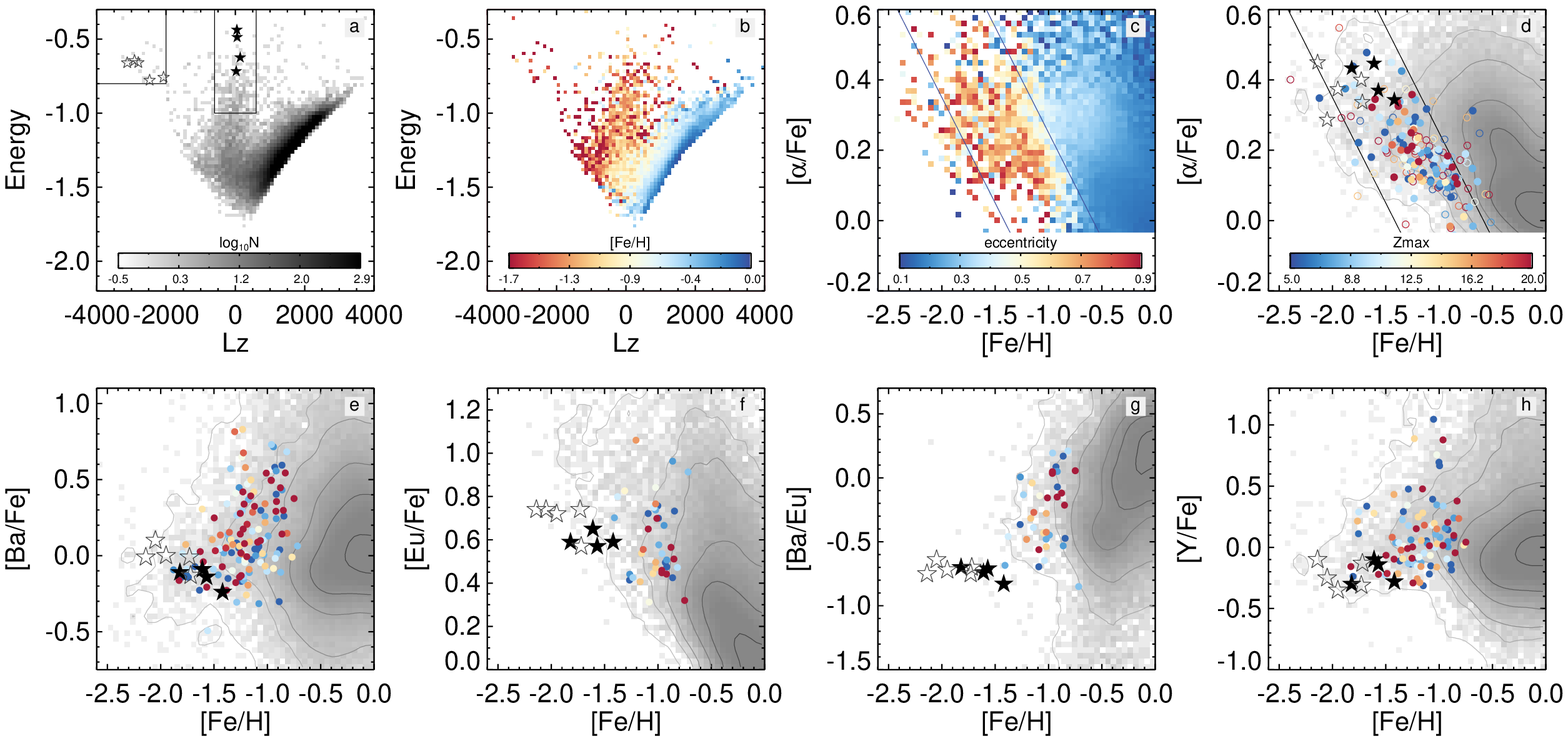}}
{\includegraphics[width=\textwidth,trim={ 0.0cm .0cm 0.cm 0cm},clip]{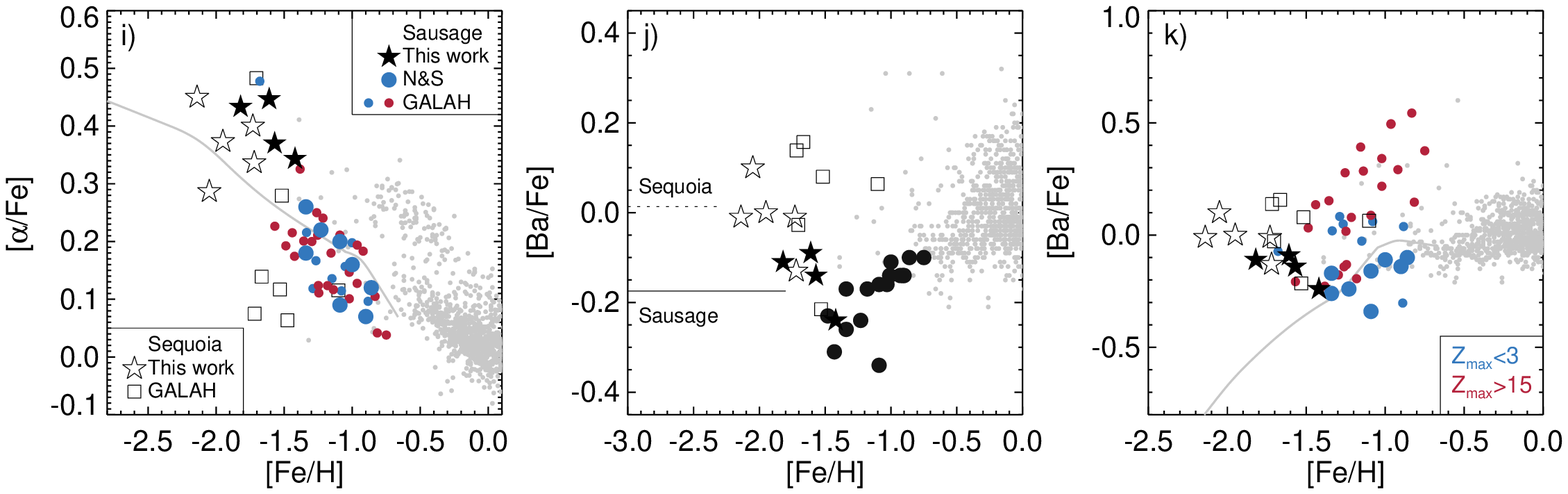}}
\end{center}
\caption{GS and Sequoia stars in GALAH DR3. {\it Top row, a:} Energy ($E$, in $10^5$ km$^2$s$^{-2}$, computed using a potential similar to \texttt{MWPotential2014} in \citet{Bovy2015} but with a virial mass of $10^{12} M_{\odot}$) and the vertical component of the angular momentum ($L_{\rm z}$, in kpc km s$^{-1}$) distribution of stars in GALAH DR3. GS (filled black stars) and Sequoia (open black stars) objects from our high-resolution campaign are also shown; these inform the placement of the selection cuts (black rectangular regions). {\it Top row, b:} $E$-$L_{\rm z}$ plane color-coded according to stellar metallicity. {\it Top row, c:} Density in the $\alpha$-[Fe/H] plane color-coded according to the eccentricity. Black diagonal lines delineate the region occupied by the GS stars. {\it Top row, d:} Greyscale density of stars in $\alpha$-[Fe/H] plane with objects from the GS box in the $E$-$L_{\rm z}$ plane overplotted; color-coding is according to the star's $Z_{\rm max}$ (highest $|Z|$ achieved by a star on its orbit). Filled-in (open) circles are stars with individual velocity component errors $<20$ km s$^{-1}$ ($<50$ km s$^{-1}$). {\it Middle row:} GALAH DR3 n-capture abundances of GS and Sequoia stars. Greyscale density shows the abundance distribution of all GALAH DR3 stars with good flags. GALAH DR3 GS stars are shown as small filled circles, color-coded according to their $Z_{\rm max}$.}
Bottom row: Combined view of GS and Sequoia. {\it Panel i:} $\alpha$ vs [Fe/H] for our high-resolution targets in GS (filled star symbols) and Sequoia (open star symbols), GALAH DR3 GS stars (blue with $Z_{\rm max}<3$ kpc and red with $Z_{\rm max}>15$ kpc small filled circles), GALAH DR3 Sequoia stars (selected using the box shown in panel {\it a} of the top row, open squares), NS GS stars (large filled blue circles) and disk stars from \citet[][]{adibekian13, delgado17} (gray dots). {\it Panel j:} Comparison of [Ba/Fe] ratio in GS and Sequoia. Mean Ba values are shown for both structures as horizontal lines. {\it Panel k:} Distinct chemo-orbital groups in GS. Note that for GALAH DR3, only stars with Ba abundance uncertainty $\le$0.1 are shown. Grey line is a Fornax dSph chemical evolution model from \citet{kobayashi20b}.
\label{fig:galah}
\end{figure*}

\section{Discussion}\label{sec:discussion}

\subsection{The r-process production in GS and Sequoia}

Fig.~\ref{elemental} compares the n-capture abundances we measure in nice stars from GS and Sequoia to those reported for dwarf spheroidals (dSphs) and ultra-faint dwarfs (UFDs), as well as for canonical Milky Way halo members from literature. Focusing on the metallicity range spanned by our high-resolution targets, a clear picture emerges. We detect an under-abundance of Sr (panel $a$, compared to halo stars) and over-abundance of Y (panel $b$, compared to e.g. Sculptor). These two so-called first-peak elements are assumed to be mostly contributed by the $s$-process, either the main one acting in low-mass AGB stars \citep{busso99} and important at high metallicities, or the weak one in massive and massive rotating stars \citep[see, e.g.,][]{kaeppeler82, pignatari10, fris12, cescu13}, dominating the low [Fe/H] range. The seeming mismatch between Sr and Y we observe is peculiar, but note that in the [Fe/H] range probed by our targets, the relative importance of the weak and main $s$-process is poorly known. However, a recent study by \citet{watson19} showed that while Sr and Y are predominantly s-process produced in the Solar System, certain level of Sr (more than Y) can be also reached by the r-process. With regards to the second-peak, we see that Ba (panel \textit{c}) is around the Solar value. Taken together, this results in a relatively low [Sr/Ba] ratio (see panel $f$ of the Figure), i.e. at the level of $-0.5<$[Sr/Ba]$<0$. As panel $f$ reveals, there are not many examples in the halo or Galactic satellites of stars with [Sr/Ba] as low as that (at a given [Ba/Fe]). Importantly, conventional $s$-process yields would predict [Sr/Ba] higher than what we observe \citep{fre10}.

The chemical enrichment pathways that shaped elemental production in the GS (and Sequoia) can be clarified considerably with the help of the Eu abundances we measure (see panel $d$ of Figure~\ref{elemental}). We detect a clear over-abundance of Eu: all our high-resolution targets are hovering around the threshold of [Eu/Fe]$\approx 0.7$ which separates the so-called r-I and r-II regimes \citep{bee05, holm20}, implying that our stars are at least moderately and perhaps strongly $r$-process enhanced. Panel $e$ demonstrates this point unambiguously: at $-2<$[Fe/H]$<-1.5$ our stars consistently display the lowest [Ba/Eu] ratios amongst the many populations studied. The prevalence of the $r$-process products would also explain the low [Sr/Ba] and possibly even [Sr/Y] ratios observed. Admittedly, our nine targets only probe the relatively metal-poor sub-population in both GS and Sequoia. In the next subsection, we use archival data to extend to higher values the metallicity range probed, going beyond the [Fe/H]$\approx-1.5$ where the $\alpha$-knee is (tentatively) located in these two systems.

\subsection{GS and Sequoia across a wide metallicity range}
 
To help build a holistic view of the abundance patterns in the GS and Sequoia, we augment our high-resolution sample with measurements from public archives, namely that of the Galactic Archaeology with Hermes (GALAH) data release 3 \citep[DR3;][]{galahdr3} and NS. GALAH DR3 and NS measurements provide an opportunity to explore the more metal-rich members of the GS and Sequoia. Appendix~\ref{sec:galah_selection} gives the details of the selection cuts applied to identify bona fide GS members amongst GALAH targets. The selection process is also illustrated in the top row of Fig.~\ref{fig:galah}. In NS, the authors performed a detailed chemical analysis of halo populations from high-resolution and high S/N UVES at VLT and FIES at NOT data. We select the likely GS members from this source following criteria similar to those applied to the GALAH data, albeit with a looser cut on energy, given much lower levels of contamination (see Appendix~\ref{sec:ns_selection}). The selection is illustrated in Fig.~\ref{fig:ns}; we take Y and Ba abundances from the NS dataset. With the inclusion of GALAH and NS data, our combined samples of GS and Sequoia stars now span a very wide range of iron abundances, namely $-2.2<\rm [Fe/H] <-0.7$.

The middle row of Fig.~\ref{fig:galah} compares n-capture element abundances of the GS and Sequoia stars in GALAH DR3 to those in our high-resolution sample. As panel $e$ of the Figure demonstrates, the GS's barium enrichment follows a clear pattern. At low metallicity $-2.2<\rm [Fe/H]<-1.5$, values of [Ba/Fe] (in fact, for both GS and Sequoia) are just around zero. For $\rm[Fe/H]>-1.5$ - as traced by GALAH - the GS [Ba/Fe] ratio exhibits a considerable bend upwards, reaching values of $\rm[Ba/Fe]\approx0.5$ and above. Note that the increasing [Ba/Fe] trend is most clear for the GS stars with high $Z_{\rm max}$ (red points), while some low-$Z_{\rm max}$ stars can be found at [Ba/Fe]$\approx0$ all the way to $\rm[Fe/H]\approx-0.7$. Next, panel $f$ of the Figure shows that with increasing metallicity, the GS stars in GALAH either stay at roughly the same level of [Eu/Fe] as reported by our high-resolution observations or drop slightly, by some 0.2\,dex. As $0.3<\rm[Eu/Fe]<1$ for the GALAH GS stars, we see that $r$-process enrichment dominates the GS chemical evolution across the entire range of metallicity. The ratio of [Ba/Eu] (shown in panel $g$) is flat for stars with $\rm[Fe/H]<-1.5$, but shows a sharp increase (or a spread) for higher iron abundances, indicating some $s$-process enrichment from AGB stars. Finally, [Y/Fe] displays a gentle slope upwards on increasing [Fe/H] as shown in panel $h$. Comparing the observed trends with models of chemical evolution \citep[see e.g.][]{Koba2020}, we hypothesise that Eu abundances reflects a switch from the pollution driven mainly by core-collapse supernova explosions at $\rm[Fe/H]<-1.5$ to increasing SN Ia contribution at higher metallicity; this can also reproduce the decrease of $\rm [\alpha/Fe]$ ratio from the same [Fe/H] in panel $i$ (see also \citet{kobayashi20b}, for the details on the SNIa and galactic chemical evolution (GCE) models). Y is contributed by AGB stars, similar to Ba, but also can be made in electron-capture supernovae \citep{Koba2020}, which may explain the Y trend that is similar (but not the same) compared to barium.

\subsection{Barium diversity in the Sausage progenitor}

The availability of the uniquely precise and pure set of measurements from NS helps us to clarify hints of the curious [Ba/Fe] behaviour revealed at $-1.5<\rm [Fe/H]<-0.7$ by the GALAH data. The GS stars we identified in the NS dataset span a very similar metallicity and [$\alpha$/Fe] range (see panel $i$ of Fig.~\ref{fig:galah}), yet [Ba/Fe] trend with increasing iron abundance is strikingly different. In the NS data, the GS stars show only moderate increase in [Ba/Fe] staying 0.1\,dex below zero all the way up to $\rm [Fe/H]=-0.7$ (see panel $j$ of the Figure). Note that, as mentioned in the Appendix~\ref{sec:ns_selection}, the bulk of the GS population mapped by NS is limited to low $Z_{\rm max}$. To investigate the dependence of barium enrichment on the stellar orbital properties, panel $k$ of Figure~\ref{fig:galah} shows GS stars in GALAH and NS split into two groups: with $Z_{\rm max}>15$ (red points) and $Z_{\rm max}<3$ (blue points). As the panel reveals, the low-$Z_{\rm max}$ stars in both GALAH and NS samples tend to have near-Solar [Ba/Fe] ratios. However, stars with high $Z_{\rm max}$ exhibit a pronounced growth in barium content to values as high as $\rm[Ba/Fe]=0.5$. 

This is a mystery since the timescales of AGB and SN Ia enrichments are similar, and the Fe production from SN Ia should suppress the [Ba/Fe] increase. Indeed, standard GCE models (see e.g. grey solid line in the bottom row of Figure~\ref{fig:galah}) can reproduce [Ba/Fe] ratios of the low  $Z_{\rm max}$ stars, but not the high $Z_{\rm max}$ stars. This is a chemical evolution model for the Fornax dSph galaxy taken from \citet{kobayashi20b}. The [$\alpha$/Fe] decrease here is mainly caused by sub-Chandrasekhar-mass SNe Ia. The metallicity of [$\alpha$/Fe] knee seems to be lower than the GS stars, which indicates that the progenitor galaxy that made the Sausage was more massive. Note that the Fornax model provides a reasonably good fit to the Sequoia's $\alpha$-sequence. The Fornax model can well reproduce the [Ba/Fe] ratios of low $Z_{\rm max}$ stars but not those of high $Z_{\rm max}$ stars. Note that at [Fe/H] $<-1.5$, the observed [Ba/Fe] ratios are clearly higher than the model; this problem is already known for the solar neighbourhood, and is due to either underestimate of the Ba production from massive stars or the ignorance of inhomogeneous enrichment from AGB stars \citep[see][for more details]{Koba2020}. Alternatively, if this barium evolution is not due to SNe Ia, then the decrease of $\rm [\alpha/Fe]$ could be explained by some other processes, e.g., low-mass core-collapse supernovae \citep{kobayashi14}. Finally, in a binary system, the old and low-mass star could be polluted with the s-process material donated by its (by now evolved into a white dwarf) companion star of intermediate mass. Such contamination may be betrayed by enhanced carbon (or nickel) abundance \citep[see, e.g.,][]{han16II}. Stars in the UVES sample do not display carbon enrichment (see Table~\ref{parameters}). Moreover, we detect no correlation between C (or Ni) and Ba in the GS stars, either in our high-resolution sample or in GALAH DR3. In principle, binarity can be identified via radial velocity oscillations or astrometric wobbling. Our cut of re-normalised unit weight error (RUWE) $<1.4$ would cull binaries with a sizes of few AUs within 1 kpc from the Sun. However, more than half of our sample lies  beyond 1 kpc, and therefore we can not completely rule out the presence of binaries at this stage.

Note that moderate variations in [Ba/Fe] ratio from increased to near-solar as a function of metallicity have been detected before in classical dwarfs \citep[][]{venncarina12} and attributed to inhomogeneous mixing of AGB products. Indeed, purely on energetic grounds, stellar envelope ejecta are predicted to mix with the interstellar medium less efficiently compared to the chemical products expelled in violent SN explosions \citep[][]{Emerick2020}. However, there may be a lot more dexterity required to explain the difference in [Ba/Fe] behaviour of the two orbital families of GS stars: a trick is needed to generate distinct amounts of Ba while living in exactly the same region of the $\alpha$-Fe plane. We note that the low-$Z_{\rm max}$ stars show a smaller dispersion in [Ba/Fe] compared to their high-$Z_{\rm max}$ counterparts. Invoking the mixing efficiency, we hypothesise that high-$Z_{\rm max}$ stars inhabited a region of the progenitor galaxy with low star formation activity, while the low-$Z_{\rm max}$ stars used to populate substantially more lively quarters.

\subsection{Sausage vs Sequoia}

Finally, let us briefly compare the chemical properties of the GS and Sequoia. We leave a thorough look at their similarities and differences to a future publication in which we will discuss all of the chemical elements available to us. Curiously, our nine high-resolution targets look rather similar in terms of their n-capture abundances despite coming from two distant corners of the $E-L_{\rm z}$ plane. Subtle differences between GS and Sequoia stars are nonetheless apparent in some of the elements discussed above. Panel $j$ of Figure~\ref{fig:galah} in particular zooms in on the [Ba/Fe] ratio in these two halo substructures. Here, we complement our measurements of the GS stars with those from NS, and add a handful of likely Sequoia members identified in the GALAH DR3. First, it is reassuring to note the agreement between our [Ba/Fe] measurements and those from the archives. Owing to the high quality of the data in hand, we note a clear offset in [Ba/Fe] between the GS and Sequoia: the Sequoia stars are $\sim$ 0.2 dex more enhanced in barium compared to the GS. Note that the difference is modest and only slightly larger than the typical measurement error ($\sim$ 0.1 dex), yet the two objects show distinct amount of barium at the same metallicity [Fe/H]$\sim-1.5$.
The $\alpha$-knees of the GS and Sequoia also show an offset, similar to what has been reported elsewhere \citep{Ma19,monty20}, and the progenitors might be more massive than Fornax.

\section{Conclusions}\label{sec:conclusion}

We present a detailed n-capture signature of four GS and five Sequoia members observed with UVES at VLT. None of these objects has been studied before using high-resolution spectroscopy. We derive accurate stellar parameters and metallicities with the FERRE code. We measure accurate Sr, Y, Ba and Eu abundances for the entire sample and report a clear and significant $r$-process enhancement. An average level of $\rm [Eu/Fe]=+0.65$ is measured for our GS and Sequoia targets with low scatter. Prominent presence of europium together with the relatively low Ba abundances ($\rm [Ba/Fe]\approx0.0$) suggest that the n-capture chemistry of both GS and Sequoia was dominated by pure $r$-process production. Additionally, we report subtle systematic differences in abundances of n-capture elements between GS and Sequoia, and point out the existence of at least two populations of GS stars with distinct stellar chemo-orbital properties, likely reflecting a range of formation conditions inside the GS progenitor galaxy.

\begin{acknowledgements}
We would like to thank Prof. Piercarlo Bonifacio for useful information of selected line list. Dr Rana Ezzeddine kindly provided high-resolution data for calibration. Dr Mat\'ias Rodr\'iguez V\'azquez helped providing theoretical and nuclear physics ideas. DA thanks the Leverhulme Trust for financial support.
C.K. acknowledges funding from the UK Science and Technology Facility Council (STFC) through grant ST/ R000905/1, and the visitor programme at Kavli Institute for Cosmology, Cambridge.

\end{acknowledgements}

\bibliography{biblio}

\appendix

\section{GALAH DR3 selection}\label{sec:galah_selection}
GALAH uses the 3.9\,m Anglo-Australian Telescope equipped with the high-resolution HERMES spectrograph covering a non contiguous 471$-$788\,nm range at resolution of 28000.  Together with stellar parameters, GALAH provides up to 30 elemental abundances including Y, Ba and Eu for a number of $\sim680,000$ nearby stars. Following recommendations from \citet{galahdr3}, we only select elemental abundances with clean flags ({$\tt flag\_x\_fe=0$)}. Note that GALAH does not cover the strongest n-capture elements absorption lines which tend to be in bluer regions. As a result, the reliability of GALAH's Y, Ba and Eu  abundance measurements decreases somewhat towards lower metallicity with only measurements at $\rm[Fe/H]>-1.5$. Top row of Figure~\ref{fig:galah} demonstrates the selection of the likely GS and Sequoia stars in the GALAH DR3 dataset. Only stars with parallax S/N$>10$ and RUWE $<1.4$ were included. Additionally, we require that for each star, the uncertainties in the individual velocity components do not exceed 50 km s$^{-1}$. As panel a in the top row reveals, the GS debris stands out clearly as a vertical column at $L_{\rm z}=0$. Accordingly, we select GS candidate stars with the following cuts: $E>-10^5$ km$^2$ s$^{-2}$ and $|L_{\rm z}|<600$ kpc km s$^{-1}$. The Sequoia candidate stars are picked using $E>-0.8 \times 10^5$ km$^2$ s$^{-2}$ and $L_{\rm z}<-2000$ kpc km s$^{-1}$. Additionally, for the GS candidate stars only, we apply a cut in the $\alpha$-[Fe/H] plane, shown in panel d of the top row of Figure~\ref{fig:galah}. This removes possible in-situ halo contamination.

\section{NS selection}\label{sec:ns_selection}
In the NS sample, the GS candidate stars are selected with the following cuts: $E>-1.25\times 10^5$ km$^2$ s$^{-2}$ and $|L_{\rm z}|<600$ kpc km s$^{-1}$. We have extended the energy threshold down compared to the cut applied for e.g. GALAH DR3 stars. This is because the NS sample consists of brighter stars with lower levels of contamination. Most of the stars are either from the GS or the Galactic in-situ halo (the Splash). Our conclusions remain unchanged if we choose a higher energy cut. Note that the absolute majority of the NS stars are on low $Z_{\rm max}$ orbits, with the bulk having $Z_{\rm max}<10$ kpc.

\begin{figure}
\begin{center}
{\includegraphics[width=\textwidth]{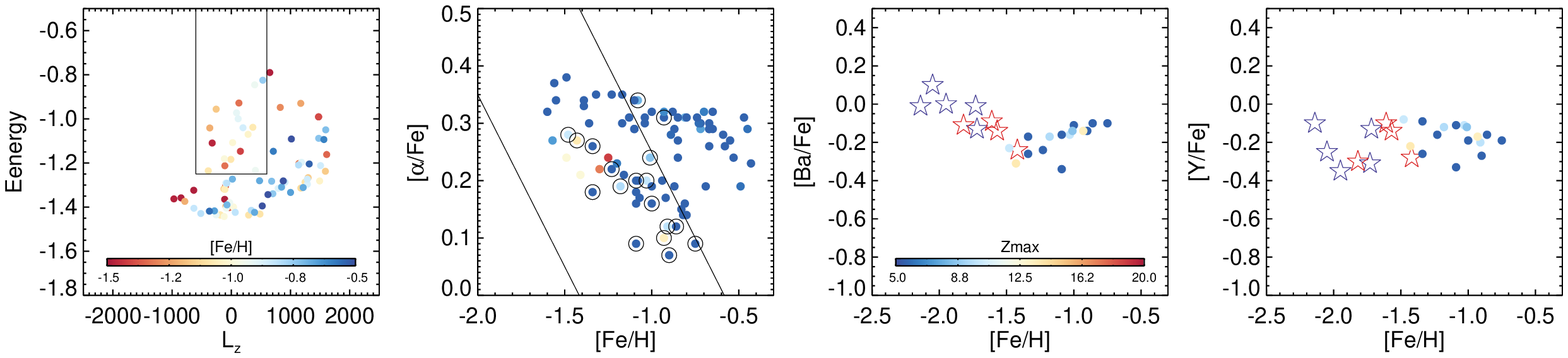}}
\end{center}
\caption{Selecting GS stars in the NS sample. {\it 1st panel (left):} Energy ($E$, in $10^5$ km$^2$s$^{-2}$, computed using a potential similar to \texttt{MWPotential2014} in \citet{Bovy2015} but with a virial mass of $10^{12} M_{\odot}$) and the vertical component of the angular momentum ($L_{\rm z}$, in kpc km s$^{-1}$) for the stars in the NS sample, color-coded according to their metallicity. {\it 2nd panel:} Distribution of NS stars in the $\alpha$-[Fe/H] plane. Points are color-coded according to their $Z_{max}$. Stars within the $E$-$L_{\rm z}$ selection box shown in the previous panel are circled. {\it 3rd panel:} [Ba/Fe] vs [Fe/H] for our high-resolution targets (GS stars in red, Sequoia stars in blue open star symbols) and the likely NS GS stars (filled circles color-coded according to their $Z_{\rm max}$) selected using the boxes shown in the 1st and the 2nd panels. {\it 4th panel:} Same as previous panel but for [Y/Fe] vs [Fe/H].}
\label{fig:ns}
\end{figure}

\end{document}

%% file: parameters2.tex
\begin{table*}
\caption{Coordinates, stellar parameters and chemical abundances. \label{parameters}}
\begin{center}
\hspace*{-2cm}\resizebox{1.1\linewidth}{!}{%
\begin{tabular}{cccccccccccc}
\hline
Star Code &  RA &      DEC &$\rm G$ & $\rm Gaia\, V_{rad}$& $\rm  V_{rad}$& $\rm t_{ exp}$&$\rm S/N$&Substructure &  $\rm T_{ eff}$ & $\log \rm g$& \\
    &$\rm deg$&$\rm deg$& mag & $\rm km\, s^{-1}$& $\rm km\, s^{-1}$  & s & 393\,nm&               &K&$\rm cm\, s^{-2}$&                \\
\hline
GY29057 &  82.21406 & $-$29.89778 & 11.5& 520.8 &521$\pm$0.3     &720 &90 & Sequoia & 5655$\pm$92 & 4.43$\pm$0.16&   \\         
GY48216 &  82.95150 & $-$36.99786 & 10.4& -43.5 &$-$43.1$\pm$0.4 &360 &66 & Sausage & 4840$\pm$48 & 1.55$\pm$0.11&    \\     
GY31607 & 104.62214 &    11.86071 &  9.7&   6.6 &6.8$\pm$0.4     &360 &82 & Sequoia & 4872$\pm$54 & 2.08$\pm$0.12&    \\      
GY57179 & 119.94508 & $-$17.38559 &  9.5& 509.7 &511.7$\pm$1.0   &360 &77 & Sausage & 4503$\pm$87 & 1.27$\pm$0.08&    \\      
GY53315 & 132.82041 & $-$44.24696 & 11.5& 619.2 &619.9$\pm$0.5   &720 &57 & Sequoia & 5136$\pm$50 & 3.95$\pm$0.12&    \\      
GY53826 & 173.57463 & $-$41.34928 & 10.9& 586.7 &586.1$\pm$0.8   &720 &77 & Sequoia & 4606$\pm$37 & 1.36$\pm$0.09&    \\      
GY61830 & 199.67778 & $-$28.84202 & 11.3& -68.5 &$-$69.4$\pm$0.5 &720 &64 & Sausage & 4971$\pm$55 & 1.69$\pm$0.12&    \\      
GY37461 & 201.87774 &    17.19132 & 11.2&  61.5 &61.9$\pm$0.7    &720 &73 & Sausage & 4516$\pm$48 & 1.56$\pm$0.07&    \\   
GY65261 & 350.43300 & $-$49.48877 &  9.9&  76.5 &77.6$\pm$0.3    &360 &80 & Sequoia & 5283$\pm$56 & 3.31$\pm$0.12&   \\
\hline\\
Gaia id&        $\left[{\rm Fe/H}\right]$ &$\left[{\rm C/Fe}\right]$ &$\left[{\rm Sr/Fe}\right]$&$\left[{\rm Y/Fe}\right]$&$\left[{\rm Ba/Fe}\right]$& $\left[{\rm Eu/Fe}\right]$&$\rm Eu_{4129}$ &$\left[{\rm \alpha/Fe}\right]^{1}$& $\left[{\rm Sr/Ba}\right]$& $\left[{\rm Ba/Eu}\right]$ \\
&&&&&&&m\AA&    &                \\
\hline
2905773322545989760&$-$2.05$\pm$0.07&$+$0.00$\pm$0.11& $-$0.35$\pm$0.10    &$-$0.25$\pm$0.09  &$+$0.10$\pm$0.06 &0.74$\pm$0.07&11.3&0.28&$-$0.45&$-$0.64  \\       
4821671436294995456&$-$1.82$\pm$0.04&$-$0.65$\pm$0.09& $-$0.51$\pm$0.10    &$-$0.30$\pm$0.11  &$-$0.11$\pm$0.08 &0.59$\pm$0.06&159.7&0.43&$-$0.40&$-$0.60   \\     
3160714468040914816&$-$1.95$\pm$0.04&$-$0.41$\pm$0.08& $-$0.51$\pm$0.09    &$-$0.35$\pm$0.10  &$+$0.00$\pm$0.06 &0.72$\pm$0.11&112.7&0.37&$-$0.51&$-$0.72   \\      
5717948445741886720&$-$1.57$\pm$0.12&$-$0.46$\pm$0.10& $-$0.44$\pm$0.10    &$-$0.14$\pm$0.10  &$-$0.14$\pm$0.09 &0.57$\pm$0.08&216.1&0.37&$-$0.30&$-$0.61   \\      
5331557897713152640&$-$2.14$\pm$0.05&$+$0.24$\pm$0.08& $-$0.12$\pm$0.08    &$-$0.10$\pm$0.09  &$-$0.01$\pm$0.07 &0.74$\pm$0.07&14.9&0.45&$-$0.11&$-$0.75   \\      
5382632652358260864&$-$1.72$\pm$0.04&$-$0.50$\pm$0.09& $-$0.33$\pm$0.10    &$-$0.13$\pm$0.12  &$-$0.13$\pm$0.08 &0.57$\pm$0.12&166.7&0.33&$-$0.20&$-$0.70   \\      
6183013242623029504&$-$1.61$\pm$0.04&$-$0.57$\pm$0.11& $-$0.39$\pm$0.09    &$-$0.10$\pm$0.10  &$-$0.09$\pm$0.11 &0.65$\pm$0.08&153.9&0.44&$-$0.30&$-$0.74   \\      
3746122603590442240&$-$1.42$\pm$0.10&$-$0.24$\pm$0.10& $-$0.14$\pm$0.14    &$-$0.28$\pm$0.09  &$-$0.24$\pm$0.07 &0.59$\pm$0.07&184.9&0.34&$+$0.10&$-$0.73   \\   
6526120553355791104&$-$1.73$\pm$0.05&$-$0.01$\pm$0.11& $-$0.31$\pm$0.09    &$-$0.31$\pm$0.09  &$-$0.01$\pm$0.07 &0.74$\pm$0.06&68.6&0.40&$-$0.30&$-$0.75  \\
\hline
\multicolumn{12}{l}{\footnotesize$^{1}\left[{\rm \alpha/Fe}\right]$ is defined for the purpose of this paper as $\left(\left[{\rm Mg/Fe}\right]+\left[{\rm Ca/Fe}\right]+\left[{\rm Ti/Fe}\right]\right)/3$.}
\end{tabular}}
\end{center}
\end{table*}

%% file: ms.bbl
\begin{thebibliography}{}
\expandafter\ifx\csname natexlab\endcsname\relax\def\natexlab#1{#1}\fi
\providecommand{\url}[1]{\href{#1}{#1}}

\bibitem[{{Abohalima} \& {Frebel}(2018)}]{jina18}
{Abohalima}, A., \& {Frebel}, A. 2018, \apjs, 238, 36

\bibitem[{{Adibekyan} {et~al.}(2013){Adibekyan}, {Figueira}, {Santos},
  {Hakobyan}, {Sousa}, {Pace}, {Delgado Mena}, {Robin}, {Israelian}, \&
  {Gonz{\'a}lez Hern{\'a}ndez}}]{adibekian13}
{Adibekyan}, V.~Z., {Figueira}, P., {Santos}, N.~C., {et~al.} 2013, \aap, 554,
  A44

\bibitem[{{Aguado} {et~al.}(2021){Aguado}, {Myeong}, {Belokurov}, {Evans},
  {Koposov}, {Allende Prieto}, {Lanfranchi}, {Matteucci}, {Shetrone},
  {Sbordone}, {Navarrete}, {Gonz{\'a}lez Hern{\'a}ndez}, {Chanam{\'e}},
  {Peralta de Arriba}, \& {Yuan}}]{agu20}
{Aguado}, D.~S., {Myeong}, G.~C., {Belokurov}, V., {et~al.} 2021, \mnras, 500,
  889

\bibitem[{{Allende Prieto} {et~al.}(2006){Allende Prieto}, {Beers}, {Wilhelm},
  {Newberg}, {Rockosi}, {Yanny}, \& {Lee}}]{alle06}
{Allende Prieto}, C., {Beers}, T.~C., {Wilhelm}, R., {et~al.} 2006, \apj, 636,
  804

\bibitem[{{Anders} {et~al.}(2019){Anders}, {Khalatyan}, {Chiappini}, {Queiroz},
  {Santiago}, {Jordi}, {Girardi}, {Brown}, {Matijevi{\v{c}}}, {Monari},
  {Cantat-Gaudin}, {Weiler}, {Khan}, {Miglio}, {Carrillo}, {Romero-G{\'o}mez},
  {Minchev}, {de Jong}, {Antoja}, {Ramos}, {Steinmetz}, \& {Enke}}]{anders2019}
{Anders}, F., {Khalatyan}, A., {Chiappini}, C., {et~al.} 2019, \aap, 628, A94

\bibitem[{{Barklem} {et~al.}(2005){Barklem}, {Christlieb}, {Beers}, {Hill},
  {Bessell}, {Holmberg}, {Marsteller}, {Rossi}, {Zickgraf}, \&
  {Reimers}}]{barklem05}
{Barklem}, P.~S., {Christlieb}, N., {Beers}, T.~C., {et~al.} 2005, \aap, 439,
  129

\bibitem[{{Beers} \& {Christlieb}(2005)}]{bee05}
{Beers}, T.~C., \& {Christlieb}, N. 2005, Highlights of Astronomy, 13, 579

\bibitem[{{Belokurov} {et~al.}(2018){Belokurov}, {Erkal}, {Evans}, {Koposov},
  \& {Deason}}]{belo18}
{Belokurov}, V., {Erkal}, D., {Evans}, N.~W., {Koposov}, S.~E., \& {Deason},
  A.~J. 2018, \mnras, 478, 611

\bibitem[{{Bensby} {et~al.}(2014){Bensby}, {Feltzing}, \& {Oey}}]{bens14}
{Bensby}, T., {Feltzing}, S., \& {Oey}, M.~S. 2014, \aap, 562, A71

\bibitem[{{Bisterzo} {et~al.}(2014){Bisterzo}, {Travaglio}, {Gallino},
  {Wiescher}, \& {K{\"a}ppeler}}]{bis14}
{Bisterzo}, S., {Travaglio}, C., {Gallino}, R., {Wiescher}, M., \&
  {K{\"a}ppeler}, F. 2014, \apj, 787, 10

\bibitem[{{Bovy}(2015)}]{Bovy2015}
{Bovy}, J. 2015, \apjs, 216, 29

\bibitem[{{Brook} {et~al.}(2003){Brook}, {Kawata}, {Gibson}, \&
  {Flynn}}]{brook03}
{Brook}, C.~B., {Kawata}, D., {Gibson}, B.~K., \& {Flynn}, C. 2003, \apjl, 585,
  L125

\bibitem[{{Buder} {et~al.}(2020){Buder}, {Sharma}, {Kos}, {Amarsi},
  {Nordlander}, {Lind}, {Martell}, {Asplund}, {Bland-Hawthorn}, {Casey}, {De
  Silva}, {D'Orazi}, {Freeman}, {Hayden}, {Lewis}, {Lin}, {Schlesinger},
  {Simpson}, {Stello}, {Zucker}, {Zwitter}, {Beeson}, {Buck}, {Casagrand e},
  {Clark}, {Cotar}, {Da Costa}, {de Grijs}, {Feuillet}, {Horner}, {Khanna},
  {Kafle}, {Liu}, {Montet}, {Nandakumar}, {Nataf}, {Ness}, {Spina}, {Traven},
  {Trepper-Garcia}, {Ting}, {Vogrincic}, {Wittenmyer}, {Zerjal}, \& {the GALAH
  collaboration}}]{galahdr3}
{Buder}, S., {Sharma}, S., {Kos}, J., {et~al.} 2020, arXiv e-prints,
  arXiv:2011.02505

\bibitem[{{Busso} {et~al.}(1999){Busso}, {Gallino}, \& {Wasserburg}}]{busso99}
{Busso}, M., {Gallino}, R., \& {Wasserburg}, G.~J. 1999, \araa, 37, 239

\bibitem[{{Cescutti} {et~al.}(2013){Cescutti}, {Chiappini}, {Hirschi},
  {Meynet}, \& {Frischknecht}}]{cescu13}
{Cescutti}, G., {Chiappini}, C., {Hirschi}, R., {Meynet}, G., \&
  {Frischknecht}, U. 2013, \aap, 553, A51

\bibitem[{{Cohen} \& {Huang}(2009)}]{cohendraco09}
{Cohen}, J.~G., \& {Huang}, W. 2009, \apj, 701, 1053

\bibitem[{{Cropper} {et~al.}(2018){Cropper}, {Katz}, {Sartoretti}, {Prusti},
  {de Bruijne}, {Chassat}, {Charvet}, {Boyadjian}, {Perryman}, {Sarri}, {Gare},
  {Erdmann}, {Munari}, {Zwitter}, {Wilkinson}, {Arenou}, {Vallenari},
  {G{\'o}mez}, {Panuzzo}, {Seabroke}, {Allende Prieto}, {Benson}, {Marchal},
  {Huckle}, {Smith}, {Dolding}, {Jan{\ss}en}, {Viala}, {Blomme}, {Baker},
  {Boudreault}, {Crifo}, {Soubiran}, {Fr{\'e}mat}, {Jasniewicz}, {Guerrier},
  {Guy}, {Turon}, {Jean-Antoine-Piccolo}, {Th{\'e}venin}, {David}, {Gosset}, \&
  {Damerdji}}]{gaia_rvs}
{Cropper}, M., {Katz}, D., {Sartoretti}, P., {et~al.} 2018, \aap, 616, A5

\bibitem[{{Dekker} {et~al.}(2000){Dekker}, {D'Odorico}, {Kaufer}, {Delabre}, \&
  {Kotzlowski}}]{dek00}
{Dekker}, H., {D'Odorico}, S., {Kaufer}, A., {Delabre}, B., \& {Kotzlowski}, H.
  2000, in \procspie, Vol. 4008, Optical and IR Telescope Instrumentation and
  Detectors, ed. M.~{Iye} \& A.~F. {Moorwood}, 534--545

\bibitem[{{Delgado Mena} {et~al.}(2017){Delgado Mena}, {Tsantaki}, {Adibekyan},
  {Sousa}, {Santos}, {Gonz{\'a}lez Hern{\'a}ndez}, \& {Israelian}}]{delgado17}
{Delgado Mena}, E., {Tsantaki}, M., {Adibekyan}, V.~Z., {et~al.} 2017, \aap,
  606, A94

\bibitem[{{Emerick} {et~al.}(2020){Emerick}, {Bryan}, \& {Mac
  Low}}]{Emerick2020}
{Emerick}, A., {Bryan}, G.~L., \& {Mac Low}, M.-M. 2020, arXiv e-prints,
  arXiv:2007.03702

\bibitem[{{Evans}(2020)}]{Ev20}
{Evans}, N.~W. 2020, in Galactic Dynamics in the Era of Large Surveys, ed.
  M.~{Valluri} \& J.~A. {Sellwood}, Vol. 353, 113--120

\bibitem[{{Ezzeddine} {et~al.}(2020){Ezzeddine}, {Rasmussen}, {Frebel},
  {Chiti}, {Hinojisa}, {Placco}, {Ji}, {Beers}, {Hansen}, {Roederer}, {Sakari},
  \& {Melendez}}]{ezzeddine20}
{Ezzeddine}, R., {Rasmussen}, K., {Frebel}, A., {et~al.} 2020, \apj, 898, 150

\bibitem[{{Frebel} {et~al.}(2010{\natexlab{a}}){Frebel}, {Kirby}, \&
  {Simon}}]{fre10}
{Frebel}, A., {Kirby}, E.~N., \& {Simon}, J.~D. 2010{\natexlab{a}}, \nat, 464,
  72

\bibitem[{{Frebel} {et~al.}(2010{\natexlab{b}}){Frebel}, {Simon}, {Geha}, \&
  {Willman}}]{freursa10}
{Frebel}, A., {Simon}, J.~D., {Geha}, M., \& {Willman}, B. 2010{\natexlab{b}},
  \apj, 708, 560

\bibitem[{{Frebel} {et~al.}(2014){Frebel}, {Simon}, \& {Kirby}}]{fresegue14}
{Frebel}, A., {Simon}, J.~D., \& {Kirby}, E.~N. 2014, \apj, 786, 74

\bibitem[{{Frischknecht} {et~al.}(2012){Frischknecht}, {Hirschi}, \&
  {Thielemann}}]{fris12}
{Frischknecht}, U., {Hirschi}, R., \& {Thielemann}, F.~K. 2012, \aap, 538, L2

\bibitem[{{Gaia Collaboration} {et~al.}(2018){Gaia Collaboration}, {Brown},
  {Vallenari}, {Prusti}, {de Bruijne}, {Babusiaux}, {Bailer-Jones}, {Biermann},
  {Evans}, {Eyer}, {Jansen}, {Jordi}, {Klioner}, {Lammers}, {Lindegren},
  {Luri}, {Mignard}, {Panem}, {Pourbaix}, {Randich}, {Sartoretti}, {Siddiqui},
  {Soubiran}, {van Leeuwen}, {Walton}, {Arenou}, {Bastian}, {Cropper},
  {Drimmel}, {Katz}, {Lattanzi}, {Bakker}, {Cacciari}, {Casta{\~n}eda},
  {Chaoul}, {Cheek}, {De Angeli}, {Fabricius}, {Guerra}, {Holl}, {Masana},
  {Messineo}, {Mowlavi}, {Nienartowicz}, {Panuzzo}, {Portell}, {Riello},
  {Seabroke}, {Tanga}, {Th{\'e}venin}, {Gracia-Abril}, {Comoretto},
  {Garcia-Reinaldos}, {Teyssier}, {Altmann}, {Andrae}, {Audard},
  {Bellas-Velidis}, {Benson}, {Berthier}, {Blomme}, {Burgess}, {Busso},
  {Carry}, {Cellino}, {Clementini}, {Clotet}, {Creevey}, {Davidson}, {De
  Ridder}, {Delchambre}, {Dell'Oro}, {Ducourant}, {Fern{\'a}ndez-
  Hern{\'a}ndez}, {Fouesneau}, {Fr{\'e}mat}, {Galluccio}, {Garc{\'\i}a-Torres},
  {Gonz{\'a}lez-N{\'u}{\~n}ez}, {Gonz{\'a}lez-Vidal}, {Gosset}, {Guy},
  {Halbwachs}, {Hambly}, {Harrison}, {Hern{\'a}ndez}, {Hestroffer}, {Hodgkin},
  {Hutton}, {Jasniewicz}, {Jean-Antoine-Piccolo}, {Jordan}, {Korn},
  {Krone-Martins}, {Lanzafame}, {Lebzelter}, {L{\"o}ffler}, {Manteiga},
  {Marrese}, {Mart{\'\i}n-Fleitas}, {Moitinho}, {Mora}, {Muinonen}, {Osinde},
  {Pancino}, {Pauwels}, {Petit}, {Recio-Blanco}, {Richards}, {Rimoldini},
  {Robin}, {Sarro}, {Siopis}, {Smith}, {Sozzetti}, {S{\"u}veges}, {Torra}, {van
  Reeven}, {Abbas}, {Abreu Aramburu}, {Accart}, {Aerts}, {Altavilla},
  {{\'A}lvarez}, {Alvarez}, {Alves}, {Anderson}, {Andrei}, {Anglada Varela},
  {Antiche}, {Antoja}, {Arcay}, {Astraatmadja}, {Bach}, {Baker},
  {Balaguer-N{\'u}{\~n}ez}, {Balm}, {Barache}, {Barata}, {Barbato}, {Barblan},
  {Barklem}, {Barrado}, {Barros}, {Barstow}, {Bartholom{\'e} Mu{\~n}oz},
  {Bassilana}, {Becciani}, {Bellazzini}, {Berihuete}, {Bertone}, {Bianchi},
  {Bienaym{\'e}}, {Blanco-Cuaresma}, {Boch}, {Boeche}, {Bombrun}, {Borrachero},
  {Bossini}, {Bouquillon}, {Bourda}, {Bragaglia}, {Bramante}, {Breddels},
  {Bressan}, {Brouillet}, {Br{\"u}semeister}, {Brugaletta}, {Bucciarelli},
  {Burlacu}, {Busonero}, {Butkevich}, {Buzzi}, {Caffau}, {Cancelliere},
  {Cannizzaro}, {Cantat-Gaudin}, {Carballo}, {Carlucci}, {Carrasco},
  {Casamiquela}, {Castellani}, {Castro-Ginard}, {Charlot}, {Chemin},
  {Chiavassa}, {Cocozza}, {Costigan}, {Cowell}, {Crifo}, {Crosta}, {Crowley},
  {Cuypers}, {Dafonte}, {Damerdji}, {Dapergolas}, {David}, {David}, {de
  Laverny}, {De Luise}, {De March}, {de Martino}, {de Souza}, {de Torres},
  {Debosscher}, {del Pozo}, {Delbo}, {Delgado}, {Delgado}, {Di Matteo},
  {Diakite}, {Diener}, {Distefano}, {Dolding}, {Drazinos}, {Dur{\'a}n},
  {Edvardsson}, {Enke}, {Eriksson}, {Esquej}, {Eynard Bontemps}, {Fabre},
  {Fabrizio}, {Faigler}, {Falc{\~a}o}, {Farr{\`a}s Casas}, {Federici},
  {Fedorets}, {Fernique}, {Figueras}, {Filippi}, {Findeisen}, {Fonti},
  {Fraile}, {Fraser}, {Fr{\'e}zouls}, {Gai}, {Galleti}, {Garabato},
  {Garc{\'\i}a-Sedano}, {Garofalo}, {Garralda}, {Gavel}, {Gavras}, {Gerssen},
  {Geyer}, {Giacobbe}, {Gilmore}, {Girona}, {Giuffrida}, {Glass}, {Gomes},
  {Granvik}, {Gueguen}, {Guerrier}, {Guiraud}, {Guti{\'e}rrez-S{\'a}nchez},
  {Haigron}, {Hatzidimitriou}, {Hauser}, {Haywood}, {Heiter}, {Helmi}, {Heu},
  {Hilger}, {Hobbs}, {Hofmann}, {Holland}, {Huckle}, {Hypki}, {Icardi},
  {Jan{\ss}en}, {Jevardat de Fombelle}, {Jonker}, {Juh{\'a}sz}, {Julbe},
  {Karampelas}, {Kewley}, {Klar}, {Kochoska}, {Kohley}, {Kolenberg},
  {Kontizas}, {Kontizas}, {Koposov}, {Kordopatis}, {Kostrzewa-Rutkowska},
  {Koubsky}, {Lambert}, {Lanza}, {Lasne}, {Lavigne}, {Le Fustec}, {Le
  Poncin-Lafitte}, {Lebreton}, {Leccia}, {Leclerc}, {Lecoeur-Taibi},
  {Lenhardt}, {Leroux}, {Liao}, {Licata}, {Lindstr{\o}m}, {Lister}, {Livanou},
  {Lobel}, {L{\'o}pez}, {Managau}, {Mann}, {Mantelet}, {Marchal}, {Marchant},
  {Marconi}, {Marinoni}, {Marschalk{\'o}}, {Marshall}, {Martino}, {Marton},
  {Mary}, {Massari}, {Matijevi{\v{c}}}, {Mazeh}, {McMillan}, {Messina},
  {Michalik}, {Millar}, {Molina}, {Molinaro}, {Moln{\'a}r}, {Montegriffo},
  {Mor}, {Morbidelli}, {Morel}, {Morris}, {Mulone}, {Muraveva}, {Musella},
  {Nelemans}, {Nicastro}, {Noval}, {O'Mullane}, {Ord{\'e}novic},
  {Ord{\'o}{\~n}ez-Blanco}, {Osborne}, {Pagani}, {Pagano}, {Pailler},
  {Palacin}, {Palaversa}, {Panahi}, {Pawlak}, {Piersimoni}, {Pineau}, {Plachy},
  {Plum}, {Poggio}, {Poujoulet}, {Pr{\v{s}}a}, {Pulone}, {Racero}, {Ragaini},
  {Rambaux}, {Ramos-Lerate}, {Regibo}, {Reyl{\'e}}, {Riclet}, {Ripepi}, {Riva},
  {Rivard}, {Rixon}, {Roegiers}, {Roelens}, {Romero-G{\'o}mez}, {Rowell},
  {Royer}, {Ruiz-Dern}, {Sadowski}, {Sagrist{\`a} Sell{\'e}s}, {Sahlmann},
  {Salgado}, {Salguero}, {Sanna}, {Santana- Ros}, {Sarasso}, {Savietto},
  {Schultheis}, {Sciacca}, {Segol}, {Segovia}, {S{\'e}gransan}, {Shih},
  {Siltala}, {Silva}, {Smart}, {Smith}, {Solano}, {Solitro}, {Sordo}, {Soria
  Nieto}, {Souchay}, {Spagna}, {Spoto}, {Stampa}, {Steele},
  {Steidelm{\"u}ller}, {Stephenson}, {Stoev}, {Suess}, {Surdej}, {Szabados},
  {Szegedi-Elek}, {Tapiador}, {Taris}, {Tauran}, {Taylor}, {Teixeira},
  {Terrett}, {Teyssandier}, {Thuillot}, {Titarenko}, {Torra Clotet}, {Turon},
  {Ulla}, {Utrilla}, {Uzzi}, {Vaillant}, {Valentini}, {Valette}, {van Elteren},
  {Van Hemelryck}, {van Leeuwen}, {Vaschetto}, {Vecchiato}, {Veljanoski},
  {Viala}, {Vicente}, {Vogt}, {von Essen}, {Voss}, {Votruba}, {Voutsinas},
  {Walmsley}, {Weiler}, {Wertz}, {Wevers}, {Wyrzykowski}, {Yoldas},
  {{\v{Z}}erjal}, {Ziaeepour}, {Zorec}, {Zschocke}, {Zucker}, {Zurbach}, \&
  {Zwitter}}]{gaia2018}
{Gaia Collaboration}, {Brown}, A.~G.~A., {Vallenari}, A., {et~al.} 2018, \aap,
  616, A1

\bibitem[{{Gallagher} {et~al.}(2020){Gallagher}, {Bergemann}, {Collet}, {Plez},
  {Leenaarts}, {Carlsson}, {Yakovleva}, \& {Belyaev}}]{gall20}
{Gallagher}, A.~J., {Bergemann}, M., {Collet}, R., {et~al.} 2020, \aap, 634,
  A55

\bibitem[{{Hanke} {et~al.}(2020){Hanke}, {Hansen}, {Ludwig}, {Cristallo},
  {McWilliam}, {Grebel}, \& {Piersanti}}]{hanke20}
{Hanke}, M., {Hansen}, C.~J., {Ludwig}, H.-G., {et~al.} 2020, \aap, 635, A104

\bibitem[{{Hansen} {et~al.}(2018{\natexlab{a}}){Hansen}, {El-Souri}, {Monaco},
  {Villanova}, {Bonifacio}, {Caffau}, \& {Sbordone}}]{hansensag18}
{Hansen}, C.~J., {El-Souri}, M., {Monaco}, L., {et~al.} 2018{\natexlab{a}},
  \apj, 855, 83

\bibitem[{{Hansen} {et~al.}(2016){Hansen}, {Andersen}, {Nordstr{\"o}m},
  {Beers}, {Placco}, {Yoon}, \& {Buchhave}}]{han16II}
{Hansen}, T.~T., {Andersen}, J., {Nordstr{\"o}m}, B., {et~al.} 2016, \aap, 588,
  A3

\bibitem[{{Hansen} {et~al.}(2018{\natexlab{b}}){Hansen}, {Holmbeck}, {Beers},
  {Placco}, {Roederer}, {Frebel}, {Sakari}, {Simon}, \& {Thompson}}]{han18}
{Hansen}, T.~T., {Holmbeck}, E.~M., {Beers}, T.~C., {et~al.}
  2018{\natexlab{b}}, \apj, 858, 92

\bibitem[{{Haywood} {et~al.}(2018){Haywood}, {Di Matteo}, {Lehnert}, {Snaith},
  {Khoperskov}, \& {G{\'o}mez}}]{Ha18}
{Haywood}, M., {Di Matteo}, P., {Lehnert}, M.~D., {et~al.} 2018, \apj, 863, 113

\bibitem[{{Helmi} {et~al.}(2018){Helmi}, {Babusiaux}, {Koppelman}, {Massari},
  {Veljanoski}, \& {Brown}}]{Helmi2018}
{Helmi}, A., {Babusiaux}, C., {Koppelman}, H.~H., {et~al.} 2018, \nat, 563, 85

\bibitem[{{Hill} {et~al.}(2002){Hill}, {Plez}, {Cayrel}, {Beers},
  {Nordstr{\"o}m}, {Andersen}, {Spite}, {Spite}, {Barbuy}, {Bonifacio},
  {Depagne}, {Fran{\c{c}}ois}, \& {Primas}}]{hill02}
{Hill}, V., {Plez}, B., {Cayrel}, R., {et~al.} 2002, \aap, 387, 560

\bibitem[{{Hill} {et~al.}(2019){Hill}, {Sk{\'u}lad{\'o}ttir}, {Tolstoy},
  {Venn}, {Shetrone}, {Jablonka}, {Primas}, {Battaglia}, {de Boer},
  {Fran{\c{c}}ois}, {Helmi}, {Kaufer}, {Letarte}, {Starkenburg}, \&
  {Spite}}]{hill19}
{Hill}, V., {Sk{\'u}lad{\'o}ttir}, {\'A}., {Tolstoy}, E., {et~al.} 2019, \aap,
  626, A15

\bibitem[{{Holmbeck} {et~al.}(2020){Holmbeck}, {Hansen}, {Beers}, {Placco},
  {Whitten}, {Rasmussen}, {Roederer}, {Ezzeddine}, {Sakari}, {Frebel}, {Drout},
  {Simon}, {Thompson}, {Bland-Hawthorn}, {Gibson}, {Grebel}, {Kordopatis},
  {Kunder}, {Mel{\'e}ndez}, {Navarro}, {Reid}, {Seabroke}, {Steinmetz},
  {Watson}, \& {Wyse}}]{holm20}
{Holmbeck}, E.~M., {Hansen}, T.~T., {Beers}, T.~C., {et~al.} 2020, \apjs, 249,
  30

\bibitem[{{Jablonka} {et~al.}(2015){Jablonka}, {North}, {Mashonkina}, {Hill},
  {Revaz}, {Shetrone}, {Starkenburg}, {Irwin}, {Tolstoy}, {Battaglia}, {Venn},
  {Helmi}, {Primas}, \& {Fran{\c{c}}ois}}]{jabsculp15}
{Jablonka}, P., {North}, P., {Mashonkina}, L., {et~al.} 2015, \aap, 583, A67

\bibitem[{{Ji} {et~al.}(2016){Ji}, {Frebel}, {Simon}, \& {Chiti}}]{jireti16}
{Ji}, A.~P., {Frebel}, A., {Simon}, J.~D., \& {Chiti}, A. 2016, \apj, 830, 93

\bibitem[{{Kaeppeler} {et~al.}(1982){Kaeppeler}, {Beer}, {Wisshak}, {Clayton},
  {Macklin}, \& {Ward}}]{kaeppeler82}
{Kaeppeler}, F., {Beer}, H., {Wisshak}, K., {et~al.} 1982, \apj, 257, 821

\bibitem[{{Kobayashi} {et~al.}(2014){Kobayashi}, {Ishigaki}, {Tominaga}, \&
  {Nomoto}}]{kobayashi14}
{Kobayashi}, C., {Ishigaki}, M.~N., {Tominaga}, N., \& {Nomoto}, K. 2014,
  \apjl, 785, L5

\bibitem[{{Kobayashi} {et~al.}(2020{\natexlab{a}}){Kobayashi}, {Karakas}, \&
  {Lugaro}}]{Koba2020}
{Kobayashi}, C., {Karakas}, A.~I., \& {Lugaro}, M. 2020{\natexlab{a}}, \apj,
  900, 179

\bibitem[{{Kobayashi} {et~al.}(2020{\natexlab{b}}){Kobayashi}, {Leung}, \&
  {Nomoto}}]{kobayashi20b}
{Kobayashi}, C., {Leung}, S.-C., \& {Nomoto}, K. 2020{\natexlab{b}}, \apj, 895,
  138

\bibitem[{{Kurucz}(2005)}]{kur05}
{Kurucz}, R.~L. 2005, Memorie della Societa Astronomica Italiana Supplementi,
  8, 14

\bibitem[{{Lai} {et~al.}(2011){Lai}, {Lee}, {Bolte}, {Lucatello}, {Beers},
  {Johnson}, {Sivarani}, \& {Rockosi}}]{laibootes11}
{Lai}, D.~K., {Lee}, Y.~S., {Bolte}, M., {et~al.} 2011, \apj, 738, 51

\bibitem[{{Lawler} {et~al.}(2001){Lawler}, {Wickliffe}, {den Hartog}, \&
  {Sneden}}]{lawler2001}
{Lawler}, J.~E., {Wickliffe}, M.~E., {den Hartog}, E.~A., \& {Sneden}, C. 2001,
  \apj, 563, 1075

\bibitem[{{Letarte} {et~al.}(2018){Letarte}, {Hill}, {Tolstoy}, {Jablonka},
  {Shetrone}, {Venn}, {Spite}, {Irwin}, {Battaglia}, {Helmi}, {Primas},
  {Fran{\c{c}}ois}, {Kaufer}, {Szeifert}, {Arimoto}, \&
  {Sadakane}}]{letfornax18}
{Letarte}, B., {Hill}, V., {Tolstoy}, E., {et~al.} 2018, \aap, 613, C1

\bibitem[{{Limberg} {et~al.}(2020){Limberg}, {Rossi}, {Beers}, {Perottoni},
  {P{\'e}rez-Villegas}, {Santucci}, {Abuchaim}, {Placco}, {Lee}, {Christlieb},
  {Norris}, {Bessell}, {Ryan}, {Wilhelm}, {Rhee}, \& {Frebel}}]{limberg20}
{Limberg}, G., {Rossi}, S., {Beers}, T.~C., {et~al.} 2020, arXiv e-prints,
  arXiv:2011.08305

\bibitem[{{Lodders}(2003)}]{lod03}
{Lodders}, K. 2003, \apj, 591, 1220

\bibitem[{{Mashonkina} \& {Christlieb}(2014)}]{mash14}
{Mashonkina}, L., \& {Christlieb}, N. 2014, \aap, 565, A123

\bibitem[{{Matsuno} {et~al.}(2019){Matsuno}, {Aoki}, \& {Suda}}]{Ma19}
{Matsuno}, T., {Aoki}, W., \& {Suda}, T. 2019, \apjl, 874, L35

\bibitem[{{Matsuno} {et~al.}(2020){Matsuno}, {Aoki}, {Casagrande}, {Ishigaki},
  {Shi}, {Takata}, {Xiang}, {Yong}, {Li}, {Suda}, {Xing}, \&
  {Zhao}}]{matsuno20}
{Matsuno}, T., {Aoki}, W., {Casagrande}, L., {et~al.} 2020, arXiv e-prints,
  arXiv:2006.03619

\bibitem[{{McMillan}(2017)}]{mcmillan2017}
{McMillan}, P.~J. 2017, \mnras, 465, 76

\bibitem[{{M{\'e}sz{\'a}ros} {et~al.}(2012){M{\'e}sz{\'a}ros}, {Allende
  Prieto}, {Edvardsson}, {Castelli}, {Garc{\'{\i}}a P{\'e}rez}, {Gustafsson},
  {Majewski}, {Plez}, {Schiavon}, {Shetrone}, \& {de Vicente}}]{mez12}
{M{\'e}sz{\'a}ros}, S., {Allende Prieto}, C., {Edvardsson}, B., {et~al.} 2012,
  \aj, 144, 120

\bibitem[{{Monty} {et~al.}(2020){Monty}, {Venn}, {Lane}, {Lokhorst}, \&
  {Yong}}]{monty20}
{Monty}, S., {Venn}, K.~A., {Lane}, J. M.~M., {Lokhorst}, D., \& {Yong}, D.
  2020, \mnras, 497, 1236

\bibitem[{{Myeong} {et~al.}(2018){Myeong}, {Evans}, {Belokurov}, {Sanders}, \&
  {Koposov}}]{mye18a}
{Myeong}, G.~C., {Evans}, N.~W., {Belokurov}, V., {Sanders}, J.~L., \&
  {Koposov}, S.~E. 2018, \apjl, 856, L26

\bibitem[{{Myeong} {et~al.}(2019){Myeong}, {Vasiliev}, {Iorio}, {Evans}, \&
  {Belokurov}}]{mye19}
{Myeong}, G.~C., {Vasiliev}, E., {Iorio}, G., {Evans}, N.~W., \& {Belokurov},
  V. 2019, \mnras, 488, 1235

\bibitem[{{Naidu} {et~al.}(2020){Naidu}, {Conroy}, {Bonaca}, {Johnson}, {Ting},
  {Caldwell}, {Zaritsky}, \& {Cargile}}]{rohan20}
{Naidu}, R.~P., {Conroy}, C., {Bonaca}, A., {et~al.} 2020, \apj, 901, 48

\bibitem[{{Nissen} \& {Schuster}(2010)}]{niss10}
{Nissen}, P.~E., \& {Schuster}, W.~J. 2010, \aap, 511, L10

\bibitem[{{Nissen} \& {Schuster}(2011)}]{Nissen11}
---. 2011, \aap, 530, A15

\bibitem[{{Nissen} \& {Schuster}(2012)}]{nissen12}
---. 2012, \aap, 543, A28

\bibitem[{{Norris} {et~al.}(2010){Norris}, {Yong}, {Gilmore}, \&
  {Wyse}}]{norrbootes10}
{Norris}, J.~E., {Yong}, D., {Gilmore}, G., \& {Wyse}, R. F.~G. 2010, \apj,
  711, 350

\bibitem[{{Pignatari} {et~al.}(2010){Pignatari}, {Gallino}, {Heil}, {Wiescher},
  {K{\"a}ppeler}, {Herwig}, \& {Bisterzo}}]{pignatari10}
{Pignatari}, M., {Gallino}, R., {Heil}, M., {et~al.} 2010, \apj, 710, 1557

\bibitem[{{Roederer} {et~al.}(2010){Roederer}, {Sneden}, {Thompson}, {Preston},
  \& {Shectman}}]{Roederer2010}
{Roederer}, I.~U., {Sneden}, C., {Thompson}, I.~B., {Preston}, G.~W., \&
  {Shectman}, S.~A. 2010, \apj, 711, 573

\bibitem[{{Ryabchikova} {et~al.}(1999){Ryabchikova}, {Piskunov}, {Savanov},
  {Kupka}, \& {Malanushenko}}]{ryab99}
{Ryabchikova}, T., {Piskunov}, N., {Savanov}, I., {Kupka}, F., \&
  {Malanushenko}, V. 1999, \aap, 343, 229

\bibitem[{{Shetrone} {et~al.}(2013){Shetrone}, {Smith}, {Stanford}, {Siegel},
  \& {Bond}}]{shedraco13}
{Shetrone}, M.~D., {Smith}, G.~H., {Stanford}, L.~M., {Siegel}, M.~H., \&
  {Bond}, H.~E. 2013, \aj, 145, 123

\bibitem[{{Simon} {et~al.}(2010){Simon}, {Frebel}, {McWilliam}, {Kirby}, \&
  {Thompson}}]{simonleo10}
{Simon}, J.~D., {Frebel}, A., {McWilliam}, A., {Kirby}, E.~N., \& {Thompson},
  I.~B. 2010, \apj, 716, 446

\bibitem[{{Sneden} {et~al.}(2008){Sneden}, {Cowan}, \& {Gallino}}]{sne08}
{Sneden}, C., {Cowan}, J.~J., \& {Gallino}, R. 2008, \araa, 46, 241

\bibitem[{{Suda} {et~al.}(2008){Suda}, {Katsuta}, {Yamada}, {Suwa}, {Ishizuka},
  {Komiya}, {Sorai}, {Aikawa}, \& {Fujimoto}}]{saga08}
{Suda}, T., {Katsuta}, Y., {Yamada}, S., {et~al.} 2008, \pasj, 60, 1159

\bibitem[{{Venn} {et~al.}(2012){Venn}, {Shetrone}, {Irwin}, {Hill}, {Jablonka},
  {Tolstoy}, {Lemasle}, {Divell}, {Starkenburg}, {Letarte}, {Baldner},
  {Battaglia}, {Helmi}, {Kaufer}, \& {Primas}}]{venncarina12}
{Venn}, K.~A., {Shetrone}, M.~D., {Irwin}, M.~J., {et~al.} 2012, \apj, 751, 102

\bibitem[{{Venn} {et~al.}(2020){Venn}, {Kielty}, {Sestito}, {Starkenburg},
  {Martin}, {Aguado}, {Arentsen}, {Bonifacio}, {Caffau}, {Hill}, {Jablonka},
  {Lardo}, {Mashonkina}, {Navarro}, {Sneden}, {Thomas}, {Youakim},
  {Gonz{\'a}lez-Hern{\'a}ndez}, {S{\'a}nchez Janssen}, {Carlberg}, \&
  {Malhan}}]{ve20}
{Venn}, K.~A., {Kielty}, C.~L., {Sestito}, F., {et~al.} 2020, \mnras, 492, 3241

\bibitem[{{Watson} {et~al.}(2019){Watson}, {Hansen}, {Selsing}, {Koch},
  {Malesani}, {Andersen}, {Fynbo}, {Arcones}, {Bauswein}, {Covino}, {Grado},
  {Heintz}, {Hunt}, {Kouveliotou}, {Leloudas}, {Levan}, {Mazzali}, \&
  {Pian}}]{watson19}
{Watson}, D., {Hansen}, C.~J., {Selsing}, J., {et~al.} 2019, \nat, 574, 497

\end{thebibliography}
